\documentclass[runningheads]{llncs}

\usepackage{enumitem}
\usepackage{subcaption}
\usepackage[utf8]{inputenc}
\usepackage{enumitem}
\usepackage{setspace}
\usepackage{afterpage}  
\usepackage{subcaption}
\usepackage{xcolor}
\usepackage{url}
\usepackage{wrapfig}
\usepackage{textcomp} 
\usepackage{comment}
\usepackage[font={scriptsize}]{caption}
\usepackage{ifthen}
\usepackage{amssymb}
\usepackage{amsmath}
\usepackage{graphicx}
\usepackage[]{easy-todo} 
\usepackage{soul}
\usepackage{multirow}
\usepackage{booktabs}
\usepackage{siunitx}
\usepackage{balance}
\usepackage{rotating}
\usepackage{tabularx}
\usepackage{float}
\usepackage{amssymb}
\usepackage{makecell}
\usepackage{expl3}
\usepackage{tablefootnote}
\usepackage{pbox}
\usepackage{mwe}
\usepackage{lipsum,etoolbox}
\usepackage{array}
\usepackage{lipsum}
\usepackage{courier}
\usepackage{listings}
\usepackage[]{easy-todo}
\usepackage{longtable}
\usepackage{pdfpages}
\usepackage{array}
\usepackage{bbding}
\usepackage{hyperref}

\graphicspath{{content/figures/}}

\definecolor{dcryellow}{rgb}{0.996, 0.627, 0.059} 
\newcommand{\conditionarrow}[2]{\mathrel{\color{dcryellow} \rightarrow\bullet}}
\definecolor{dcrblue}{rgb}{0.129, 0.573, 1.000}
\newcommand{\responsearrow}[2]{\mathrel{\color{dcrblue} \bullet\rightarrow}}
\definecolor{dcrgreen}{rgb}{0.448, 0.773, 0.401}
\newcommand{\inclusionarrow}[2]{\mathrel{\color{dcrgreen} \rightarrow \! + \!}}
\definecolor{dcrred}{rgb}{0.984, 0.298, 0.298}
\newcommand{\exclusionarrow}[2]{\mathrel{\color{dcrred} \rightarrow \!\% \!}}
\definecolor{ao(english)}{rgb}{0.0, 0.5, 0.0}

\definecolor{cResponse}{HTML}{2192FF}
\definecolor{cCondition}{HTML}{FFA30D}
\definecolor{cMilestone}{HTML}{A932D0}
\definecolor{cExclude}{HTML}{FF0000}
\definecolor{cInclude}{HTML}{30A523}
\definecolor{cPreCondition}{HTML}{FFA30D}
\definecolor{cNoResponse}{HTML}{7A514D}
\definecolor{cLogicalInclude}{HTML}{30A523}
\definecolor{cSpawn}{HTML}{2B4863}

\newcolumntype{P}[1]{>{\raggedleft\arraybackslash}p{#1}}
\newcolumntype{L}[1]{>{\raggedright\arraybackslash}p{#1}}
\usepackage{floatpag}
  \usepackage{graphicx,scalerel}

\makeatletter
\def\els@aparagraph[#1]#2{\elsparagraph[#1]{#2}}
\def\els@bparagraph#1{\elsparagraph*{#1}}
\makeatother

\begin{document}
\title{Improving the Understandability of Conceptual Models via Abstract Notation Engineering}
%
%
\author{Amine Abbad-Andaloussi \and Daniel Jeppe Schütt \and Kasper Solhøj Jørgensen \and Tobias van Deurs Lundsgaard \and Hugo A. López}
\authorrunning{Abbad-Andaloussi et al.}
 \titlerunning{Understandability of Conceptual
Models via Abstract Notation Engineering}
%
\institute{}
\maketitle              
\begin{abstract}
Conceptual modeling supports the design, analysis, and communication of the properties of complex systems, yet conceptual models can be difficult to understand when domain-level abstractions must be encoded through low-level constructs required mainly for semantic conformity. Prior work has mainly improved how existing individual constructs are visually represented. We shift the focus from individual constructs to recurring configurations of constructs, and propose abstract notation engineering as a language-agnostic method for replacing such configurations with higher-level, semantically transparent constructs. The method comprises pattern identification, pattern formalization, visual notation design, and empirical validation. We instantiate it for Dynamic Condition Response (DCR) graphs, where common workflow patterns require elaborate low-level configurations. The resulting extension, DeCleaR, replaces such configurations with compact pattern-based abstractions. The results of our empirical validation show that DeCleaR improves perceived empirical quality, pragmatic quality, and user preference over standard DCR graphs.


\end{abstract}

\section{Introduction}
\label{section:introduction}

Conceptual modeling is fundamental for the design, analysis, and communication of the characteristics of complex systems~\cite{karagiannis2016domain}. Languages such as the Unified Modeling Language (UML), Entity-Relationship diagrams (ERD), Business Process Model and Notation (BPMN), and Dynamic Condition Response (DCR) enable modelers to capture static and dynamic aspects of systems, including organization, operational logic, and data constraints, with formal precision. To support understandability, these languages rely on graphical notations~\cite{dumas2013fundamentals}.

The effectiveness of conceptual models depends on how well they convey concepts and interrelations. While the semantic capabilities of modeling languages have received considerable attention~\cite{dumas2013fundamentals}, their pragmatic aspects are less understood. Indeed, conceptual modeling languages are typically optimized for semantic correctness rather than pragmatics. For example, representing a many-to-many relationship in an ER diagram requires an associative entity and multiple connecting relationships. The intention is captured, but through artificial constructs foreign to the modeler's domain,  needed mainly to satisfy the language semantics. Behavioral languages face similar issues. Translating a BPMN model into a workflow net may require auxiliary places that are absent from the original BPMN representation but needed to preserve behavior. This creates an understandability gap, as practitioners must distinguish domain-relevant constructs from helper constructs introduced for semantic conformity. From a cognitive systems perspective, modeling involves an \emph{encoding} phase, where practitioners' mental models are translated into a conceptual model, and a \emph{decoding} phase, where the conceptual model is interpreted back into mental models. The greater the distance between mental models and their conceptual representation, the more effort users are expected to exert during comprehension tasks~\cite{Zugal2013b,siegmund2017measuring}.

A major contribution to pragmatic aspects of conceptual models is Moody's Physics of Notations (PoN)~\cite{moody2009physics}, which has influenced the representation of BPMN~\cite{genon2010analysing}, DCR graphs~\cite{lopez2022re}, and other notations. A central PoN principle is \emph{semantic transparency}, denoting the degree to which a visual construct's meaning can be inferred from its appearance~\cite{moody2009physics}. Achieving semantic transparency remains challenging across languages~\cite{trinh2023semantic,moody2009physics}. Prior work has mainly improved the transparency of individual visual constructs, for instance, by redesigning constraints between process model activities with alternative glyphs~\cite{blasilli2025improving,trinh2023semantic}, exploring alternative 3D representations \cite{3DDCR}, or overlaying domain-specific representations~\cite{jensen2024towards}. However, these efforts focus on notational elements \emph{in isolation}.

This work shifts the focus from improving individual constructs to abstracting recurring configurations of constructs into higher-level, semantically transparent constructs. We refer to this approach as \emph{abstract notation engineering}. It aims to identify mental abstractions used by practitioners and represent them within an existing modeling language. The approach is inspired by earlier abstraction efforts in conceptual modeling, namely, \textit{Declare}, which can be seen as an abstraction over recurring Linear Temporal Logic (LTL) formulas in industrial settings~\cite{dwyer1999patterns}. This  initiative showed that recurring low-level configurations can be lifted into reusable modeling constructs. However, Declare was not developed as part of a systematic notation engineering method focusing on \textit{understandability}. Besides, common abstraction mechanisms mainly take the form of \emph{modularization}, where constructs are grouped into containers that hide internal details~\cite{Zugal2013b}. Such mechanisms exist in several notations, including subprocesses in BPMN~\cite{dumas2013fundamentals}, nestings~\cite{hildebrandt2011nested}, and links in DCR graphs~\cite{debois2020chain}. Although modularization reduces visible complexity, it does not necessarily introduce higher-level constructs aligning with practitioners' mental abstractions. Once the container is expanded, the same low-level configuration reappears and must still be interpreted. Abstract notation engineering, therefore, combines abstraction with \textit{encapsulation} to define \emph{behavioral encodings}~\cite{yueDCRBPMN}, i.e., language fragments that behave as a unit according to an intended higher-level concept. These encodings provide the semantic basis for new constructs, which must be complemented by \emph{symbol elicitation} to ensure that their meaning can be inferred from their appearance. The obtained visual abstractions require then \textit{empirical validation} to assess their alignment with practitioners' mental schemas and understandability.
\raggedbottom

In a nutshell, our main contribution is a language-agnostic method for enhancing the understandability of conceptual modeling notations through abstract notation engineering. We demonstrate the method on declarative process modeling languages, focusing on DCR graphs. Declarative languages, such as Declare~\cite{pesic2007declare} and DCR Graphs~\cite{hildebrandt2011declarative}, specify constraints while leaving execution order open~\cite{andaloussi2020declarative}. This flexibility is valuable for constraint-governed processes in banking, healthcare, and law, where decisions are discretionary, and the space of possible executions is too large to prescribe as sequential flows. However, it also incurs a representational cost, since common control-flow patterns, such as a sequence of two activities or an exclusive choice, require elaborate constraint configurations~\cite{fahland2009declarative}. We use DCR as a representative for declarative languages, given its consistently demonstrated practical value in digitalization initiatives over the years~\cite{keramidis2026business,hildebrandt2020ecoknow}. Applying our method to DCR resulted in \textit{\textbf{D}e\textbf{C}lea\textbf{R}}, a set of nine pattern abstractions for recurring process behaviors. We evaluated selected DeCleaR patterns empirically and found significant improvements in perceived empirical quality, perceived pragmatic quality, and user preference compared with standard DCR representations. With this contribution, we aim to support the adoption of declarative languages among industrial users of DCR, while also providing a method that other notations, such as Declare, BPMN, UML, and ER, could adopt. The remainder of the paper is structured as follows. Section~\ref{sec:background} reviews relevant background. Section~\ref{sec:method} presents our method. Section~\ref{sec:usecase} instantiates the method on DCR graphs and presents our empirical evaluation. Section~\ref{sec:discussion} discusses findings and limitations. Section~\ref{sec:conclusion} concludes the paper.

\vspace{-0.2cm}
\section{Background and Related Work}
\label{sec:background}

\subsection{Empirical Work on the Understandability of Conceptual Models}\label{sec:clt}

Empirical work on the understandability of conceptual models can be divided into studies investigating the (1) \textit{presentation}, (2) \textit{users}, and (3) \textit{tasks} in which models are used~\cite{figl2017comprehension,krogstie2012model}. This work subscribes to the first stream, which includes studies on presentation mediums and tools~\cite{turetken2016effect,abbad2019exploring}, and studies on the syntax, semantics, and visual representation of conceptual models~\cite{winter2020measuring,figl2020declarative,trinh2023semantic,abbad2025model,figl2017comprehension}.

Focusing on visual representations, prior research has examined a range of modeling languages (overview in~\cite{figl2017comprehension}). In this paper, declarative languages such as Declare and DCR graphs are particularly relevant because their declarative nature creates representational challenges despite modeling advantages and increasing  adoption~\cite{andaloussi2020declarative,blasilli2025improving,hildebrandt2020ecoknow} (cf. Sect. \ref{section:introduction}). Existing studies have proposed new visual representations and modeling guidelines to support modeling practices and model comprehension~\cite{trinh2023semantic,blasilli2025improving,3DDCR,jensen2024towards,hanser2016new}. However, these works focus on improving individual symbol representations, and no prior work has proposed an empirically guided, language-agnostic method for improving visual representations using abstraction techniques, which constitutes the core focus of this paper.

\subsection{Improving Modeling Notations}\label{subsect:improvingViz}

A substantial body of literature has examined how diagrammatic representations, including conceptual and process models, can support understandability~\cite{moody2009physics}. Two theoretical perspectives are particularly informative for this study. PoN formulates design principles for improving conceptual model comprehension, including Semiotic Clarity, Perceptual Discriminability, Semantic Transparency, Complexity Management, Cognitive Integration, Visual Expressiveness, Dual Coding, Graphic Economy, and Cognitive Fit~\cite{moody2009physics}. Relevant here are \emph{semantic transparency}, which concerns whether a symbol's meaning can be inferred from its visual appearance, and \emph{dual coding}, which concerns the complementarity of visual and textual elements~\cite{moody2009physics}.

Apart from PoN, two major frameworks consider the understandability of conceptual models. The \textit{Semiology of Graphics}~\cite{bertin1983semiology} identifies fundamental \emph{visual variables}, including shape, color, size, orientation, brightness, and texture. \textit{SEQUAL}~\cite{krogstie2012model} provides quality dimensions, including \emph{empirical quality}, which addresses how clearly and distinctly representation elements can be identified, and \emph{pragmatic quality}, which concerns whether the intended meaning is correctly understood. These foundations structure the design and evaluation of our abstract notation engineering approach in Sect.~\ref{sec:method}.

\subsection{DCR Graphs}

\begin{figure}[t]
    \vspace{-0.5cm}
    \centering
    \includegraphics[scale=0.25]{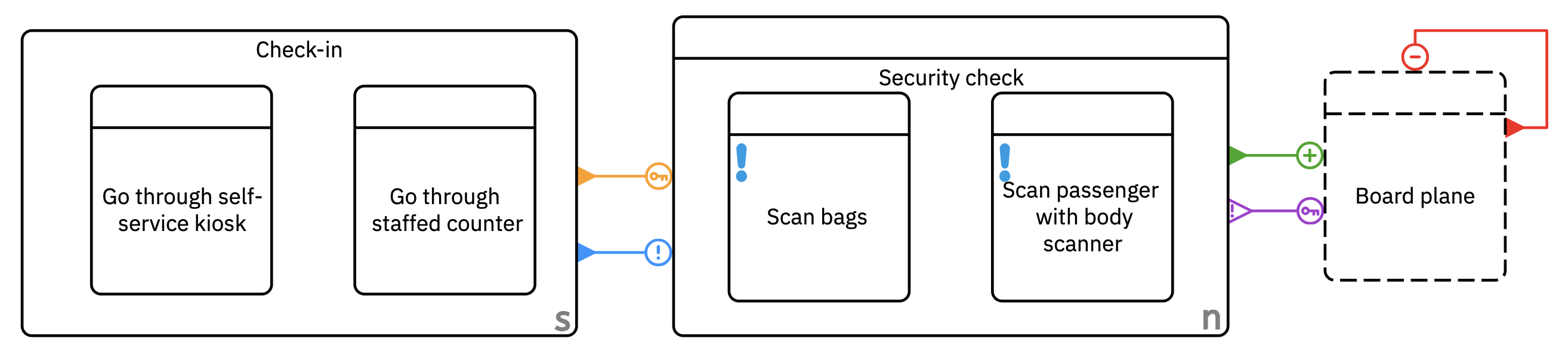} 
    \caption{Simplified passenger airport journey modeled in DCR Graphs. }
    \label{fig:dcr-example}
    \vspace{-0.5cm}
\end{figure}

A DCR graph~\cite{hildebrandt2011declarative} represents processes as a multidirected graph where nodes correspond to events (i.e., process activities) and edges define behavioral constraints. We introduce the main notation elements via Fig.~\ref{fig:dcr-example}. Each event appears as a rectangle and can have different states based on its current \textit{marking}. A blue exclamation mark denotes a pending event, i.e., an obligation that must be fulfilled before the process can complete. Events with dashed borders are excluded and temporarily unavailable for execution. A green checkmark indicates executed status, meaning the activity has occurred. 

The edges between events represent DCR constraints. We use the notational variant introduced in~\cite{trinh2023semantic}: (1) \textit{Condition} \textcolor{dcryellow}{\textbf{\textit{e}} \scalerel*{\includegraphics{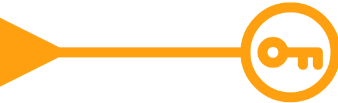}}{A} \textbf{\textit{f}}}: \textit{f} cannot be executed until \textit{e} has been executed, or \textit{e} is excluded; (2) \textit{Response} \textcolor{dcrblue}{\textbf{\textit{e}} \scalerel*{\includegraphics{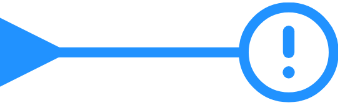}}{A} \textbf{\textit{f}}}: when \textit{e} executes, \textit{f} becomes pending and must eventually occur or be excluded; (3) \textit{Dynamic inclusion} \textcolor{ao(english)}{\textbf{\textit{e}} \scalerel*{\includegraphics{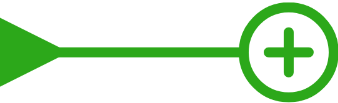}}{A} \textbf{\textit{f}}}: after executing \textit{e}, \textit{f} is included among the possible actions; (4) \textit{Dynamic exclusion} \textcolor{dcrred}{\textbf{\textit{e}} \scalerel*{\includegraphics{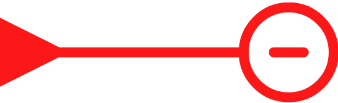}}{A} \textbf{\textit{f}}}: after executing \textit{e}, \textit{f} is excluded from the possible actions; and (5) \textit{Milestone} \textcolor{violet}{\textbf{\textit{e}} \scalerel*{\includegraphics{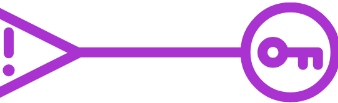}}{A} \textbf{\textit{f}}}: if \textit{e} is pending, \textit{f} cannot be enabled.

DCR uses nesting~\cite{hildebrandt2011nested} (rectangles with \textbf{n}) and subprocesses~\cite{debois2014hierarchical} (rectangles with \textbf{s}) to group related events. \emph{Nesting} reduces diagram complexity: an edge pointing to or from a nesting event behaves as if each event inside the nesting had that edge. Subprocesses add a semantic layer by endowing the outer layer (i.e., the subprocess) with an execution marking whose state depends on the markings and constraints of its internal events. Thus, a subprocess is completed once it reaches an accepting state, i.e., no pending events.

Fig.~\ref{fig:dcr-example} presents a DCR graph modeling a (simplified) airport passenger journey from check-in to boarding. The process begins with a ``Check-in'' subprocess containing two alternatives: ``Go through self-service kiosk'' and ``Go through staffed counter''. Since these events are inside a subprocess, ``Check-in'' reaches an accepting state as soon as one alternative is executed. This also means that the two outbound DCR relations take effect once one of the two ``Check-in'' events is executed. The ``Security check'' is a nest with two required events, ``Scan bags'' and ``Scan passenger with body scanner''. The condition between ``Check-in'' and ``Security check'' states that the nested security events cannot execute until ``Check-in'' has been completed at least once. The response relation specifies that completing ``Check-in'' triggers a requirement to undergo the ``Security check''. The milestone relation from ``Security Check'' to ``Board plane'' prevents boarding while security obligations remain pending. Hence, boarding becomes available only after all security requirements are completed. ``The Board Plane'' event has a self-referential exclude relation, removing boarding from the process upon execution and preventing multiple boardings. The include relation from ``Security check'' to ``Board Plane'', however, allows including ``Board Plane'' again if another security check is performed.




\section{Method}
\label{sec:method}

\begin{figure}[t]
\centering
\vspace{-0.4cm}
  \includegraphics[scale=0.4]{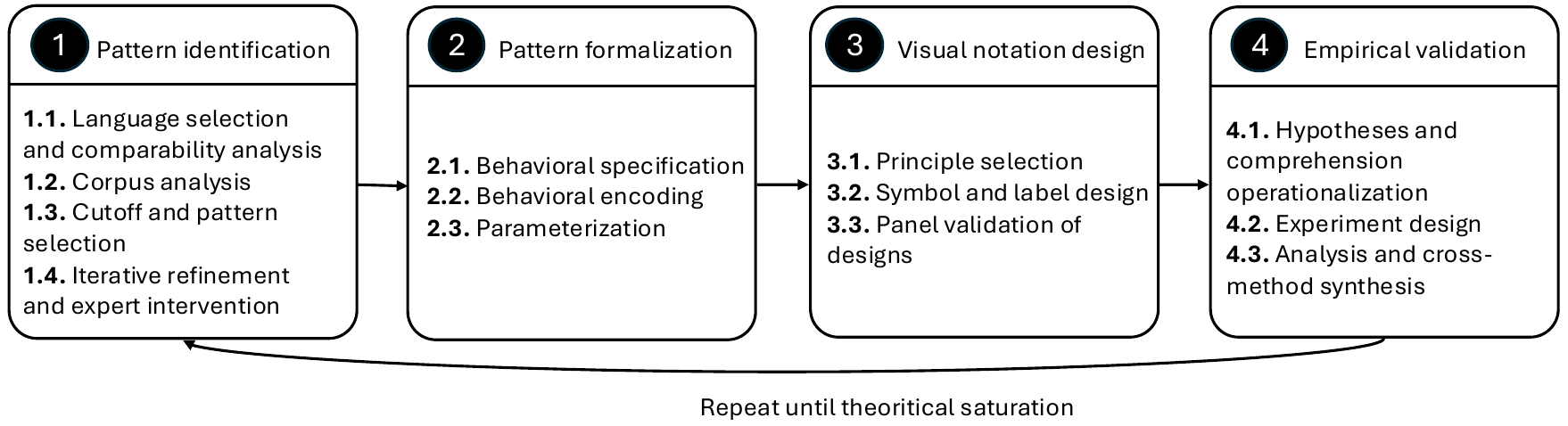} 
\caption{Overview of the notation engineering method.}
\label{fig:notation-engineering-method}
\vspace{-0.5cm}
\end{figure}

This section presents our method for designing visual notation abstractions. The method is language-agnostic and can be applied to any formal language with a visual representation and an underlying computational semantics. It represents one iteration of an iterative abstract notation engineering process: candidate patterns are identified, formalized, visually designed, and empirically evaluated. The results can inform subsequent refinements, either by redesigning existing abstractions or by adding further patterns, until theoretical saturation is reached, i.e., until further iterations reveal no substantially new patterns, design issues, or empirical insights. Its four steps build upon established foundations: \textit{Pattern identification} (Step~1) is based on pattern-oriented conceptual modeling~\cite{russell2006workflow,elgammal2016formalizing}, where recurring patterns are organized into catalogues to ease modeling practices. \textit{Pattern formalization} (Step~2) draws on test-driven modeling~\cite{christfort2025static} and behavioral encodings~\cite{yueDCRBPMN}. \textit{Visual notation design} (Step~3) is grounded in visual notation design principles~\cite{moody2009physics,bertin1983semiology}. Finally, \textit{empirical validation} (Step~4) follows empirical research guidelines~\cite{wohlin2003empirical}. The method was refined by the co-authors over six months during its application on the DCR use case (Sect.~\ref{sec:usecase}), until convergence on the current version. Fig. \ref{fig:notation-engineering-method} provides an overview of our method.


\subsection{Step 1: Pattern Identification}\label{sec:step1}

The first step determines which patterns would benefit most from pattern-based abstractions for a specific language or community. It comprises four activities.

\vspace{0.1cm}

\noindent\emph{Step~1.1: Language Selection and Compatibility Analysis.} We begin by selecting a target language and evaluating which patterns can be expressed in it. Established catalogues, such as the 43 workflow patterns~\cite{russell2006workflow}, can serve as a starting point for process-oriented languages, while domain-specific catalogues should be consulted for other languages. Each candidate pattern should be assessed by asking: \textit{Can the target language express the pattern's behavior through its existing constructs?} Inexpressible patterns should be excluded.

\vspace{0.1cm}

\noindent\emph{Step~1.2: Corpus Analysis.} Compatibility alone does not ensure relevance. Our method therefore, encourages analyzing natural language process specifications to determine how frequently each compatible pattern occurs in practice. 
Annotation studies~\cite{lopez2025ambiguity} of language patterns could reveal those that dominate daily practice and would thus yield better conceptual model comprehension if abstracted.

\vspace{0.1cm}

\noindent\emph{Step~1.3: Cutoff and Pattern Selection.} A threshold-based cutoff may be applied to the annotation results to select a concise set of patterns. 

\vspace{0.1cm}

\noindent\emph{Step~1.4: Iterative Refinement and Expert Intervention in Pattern Selection.} Pattern selection should be iterative, allowing refinement based on expert input and emerging insights. Rather than fixing the pattern set upfront, domain expertise should be incorporated throughout this step. Experts may propose additional patterns not strongly represented in the literature, yet relevant and useful for the target use case. Their inclusion is permitted if compatibility with the target modeling language is verified. We recommend balancing coverage, i.e., abstracting enough patterns to be useful, against complexity, i.e., keeping the number of abstractions manageable for users to learn.

\subsection{Step 2: Pattern Formalization}\label{sec:step2}

Each pattern must be defined in terms of the host notation's semantics, ensuring that the abstraction is a semantically grounded configuration of constructs. This step contains three activities. 

\vspace{0.1cm}

\noindent\emph{Step~2.1: Behavioral Specification.} For each pattern, the method defines its intended behavior through representative valid and invalid execution traces. This trace-based characterization specifies \emph{what} the pattern means in observable behavior before defining \emph{how} it is implemented. It may also include modeling decisions to resolve ambiguities.

\vspace{0.1cm}

\noindent\emph{Step~2.2: Behavioral Encodings.}
This step extends the compatibility analysis by translating abstracted patterns into model fragments in the host notation. Each fragment captures the abstraction's semantic content and serves as its computational counterpart. This backward mapping maintains compatibility with the original notation and ensures that existing computational support, such as execution, simulation, verification, and conformance checking, remains applicable. A key constraint is the \emph{encapsulation rule}, i.e., no relations may cross the boundary of the abstracted fragment. This keeps the abstraction self-contained and substitutable without side effects on the surrounding conceptual model.

\vspace{0.1cm}

\noindent\emph{Step~2.3: Parameterization.} Some patterns may require parameters, such as ``at least $n$ occurrences'' of an event. The formalization should capture them.

\subsection{Step 3: Visual Notation Design}\label{sec:step3}

The third step designs the visual representation of each pattern view. It comprises three activities.

\vspace{0.1cm}

\noindent\emph{Step~3.1: Principle Selection.}
The designer should select visual design principles to guide the new representation. Established frameworks include \textit{PoN}~\cite{moody2009physics} and the \textit{Semiology of Graphics}~\cite{bertin1983semiology} (cf. Sect.~\ref{subsect:improvingViz}). Other frameworks, such as Gestalt principles of perception \cite{todorovic2008gestalt}, may apply depending on the domain context. Based on the literature~\cite{moody2009physics,bertin1983semiology}, we posit that designers should prioritize two concerns: (i)~\emph{interpretability}, ensuring that each symbol's meaning can be inferred from its appearance, and (ii)~\emph{distinguishability}, ensuring that symbols for different patterns are visually distinct from one another and from existing constructs.

\vspace{0.1cm}

\noindent\emph{Step~3.2: Symbol and Label Design.} For each pattern, candidate visual constructs, i.e., symbols, and labels should be generated. A symbol should depict the behavioral essence of the pattern in a way recognizable to practitioners. The label should provide a textual anchor that names the pattern in clear, unambiguous terminology. Both should maximize interpretability and distinguishability.

\vspace{0.1cm}

\noindent\emph{Step~3.3: Panel Validation of Designs.} To ensure that the chosen symbols align with practitioners' mental schemas, our method adopts a structured elicitation procedure based on the Delphi method~\cite{okoli2004delphi}. This method gathers expert input over several iterations, each time sharing a summary of group answers so experts can review their opinions and converge toward agreement~\cite{okoli2004delphi}. Applied to our context, experts should evaluate candidate designs across multiple rounds to reach a consensus that maximizes interpretability and distinguishability.

\subsection{Step 4: Empirical Validation}\label{sec:step4}

The fourth step evaluates the chosen patterns from a user perspective. Here, we focus only on the aspects most relevant for evaluating visual notations, while comprehensive guidelines for conducting empirical studies can be found in \cite{wohlin2003empirical}.

\vspace{0.1cm}

\noindent\emph{Step~4.1: Hypotheses and Comprehension Operationalization.} Empirical evaluation may target several activities, including model comprehension, modeling, and maintenance. Because comprehension is inherent to all these activities, we recommend prioritizing it first. In this context, the independent variable should be the notation condition, i.e., abstracted versus standard notation. The dependent variables should capture separate comprehension dimensions rather than a single one, since a visual abstraction may improve users' perception of a model's clarity and interpretability without improving how accurately they decode its behavior. Relevant measures include \emph{answer correctness}, assessed through decoding tasks (e.g., producing valid execution traces for a given model~\cite{abbad2023complexity}); \emph{perceived quality}, captured with questionnaire items grounded in SEQUAL~\cite{krogstie2012model}; and \emph{preference}, quantified through comparisons between abstracted and standard notations.  Measures, such as eye tracking or  EEG, can be additionally incorporated to capture the cognitive processes underlying model comprehension~\cite{holmqvist2011}.

\vspace{0.1cm}

\noindent\emph{Step~4.2: Experimental Design.} Participants, stimuli, instruments, and procedure should follow established guidelines~\cite{wohlin2003empirical}. Both within-subject and between-subject designs could be applicable, depending on the study's objectives. A within-subject design, where each participant interprets models both with and without visual pattern abstraction, is preferable when the participant pool is small because it results in more data points for statistical analyses by using each participant as their own control and eliminates individual differences as a confounding factor. 
A between-subject design may be appropriate to avoid fatigue due to long experiments and the learning effect between conditions, common in within-subject settings.
Yet, the learning effect could also be mitigated in a within-subject design by counterbalancing the presentation order. Each pattern should be embedded in an identical base model, so the notation variant is the only manipulated variable. A mixed-methods approach is advisable. Beyond quantitative instruments, semi-structured interviews can explore how participants interpret visual elements and what challenges their understanding.

\vspace{0.1cm}

\noindent\emph{Step~4.3: Analysis and Cross-method Synthesis.} Hypotheses should be tested with statistical tests suited to the data properties~\cite{wohlin2003empirical}. In addition, the empirical study can be complemented by interviews to gain qualitative insights about users' experience. Herein, interview transcripts should undergo structured qualitative coding~\cite{CodingManualQualitative}, with independent coders and inter-rater reliability checks. Quantitative and qualitative insights should then be related to better understand the observed effects. This integration makes the empirical study a source of actionable design insights.

\section{Use Case: Instantiation in DCR Graphs}
\label{sec:usecase}

We demonstrate our method by instantiating it on DCR. The instantiation produces DeCleaR, an extension that equips DCR with nine abstracted patterns. 

\subsection{Step 1: Pattern Identification}
\label{sec:usecase-step1}



\noindent\textit{Language Selection and Compatibility Analysis (Step 1.1). }
DCR was selected as the target language owing to the challenges practitioners face when expressing common flow-based patterns under its declarative, constraint-based semantics. While compatibility analyses of the 43 workflow patterns of Russell et al.~\cite{russell2006workflow} exist for BPMN~\cite{zeising2014towards}, YAWL~\cite{han2012control}, and CMMN~\cite{carvalho2016analysis}, no such analysis had been undertaken for DCR. We therefore examined the 43 patterns for compatibility with DCR constructs and documented how, or whether, each behavior can be reproduced using combinations of DCR constructs. Patterns that could not be faithfully expressed within the basic DCR formalism\footnote{By ``basic'' we refer to whether the patterns could be encoded in the original presentation of DCR graphs~\cite{hildebrandt2011declarative}. Some patterns could be expressed in extensions of DCR graphs with time or data, but they were deemed out of scope in this iteration.} were excluded. {The full results are available in our online repository~\cite{appendix}}.
\vspace{0.1cm}

\noindent\textit{Corpus Analysis  (Step 1.2).}
To determine which expressible patterns appear frequently enough to justify abstraction, we analyzed 50 declarative process descriptions from a study on textual ambiguity in business process models~\cite{lopez2025ambiguity}. The analysis followed a sentence-level scanning procedure. For each sentence, we determined whether it fully instantiated one compatible workflow pattern and incremented the corresponding occurrence count. Partial matches were carried forward and checked against the subsequent sentence before any tally decision. {The complete annotation results are available in our online repository  \cite{appendix}}.
\vspace{0.1cm}

\noindent\textit{Cutoff and Pattern Selection  (Step 1.3).} 
The analysis produced a clear frequency stratification. The top seven patterns were observed 84, 46, 24, 23, 21, 20, and 18 times, respectively, across the 50 descriptions. The eighth-ranked pattern appeared only 7 times, a sharp drop that marked a natural boundary between widely-used and rarely-used patterns.
\vspace{0.1cm}

\noindent\emph{Iterative Refinement and Expert Intervention in Pattern Selection (Step~1.4).} 
Given the iterative nature of our approach, we also explored compliance patterns~\cite{elgammal2016formalizing}. Although less widely adopted than Russell’s patterns~\cite{russell2006workflow}, they capture compliance modeling behaviors relevant to the DCR community and compliant-by-design Case Management Systems~\cite{hildebrandt2020ecoknow}. Applying the same compatibility procedure confirmed that six compliance patterns could be expressed in DCR, expanding the candidate pool to thirteen patterns. Following co-author discussions guided by the industry-oriented DCR expertise of some of them, nine patterns were selected by balancing coverage against complexity. These became the final DeCleaR pattern set. Table~\ref{table:patterns} lists the nine identified patterns with concise descriptions.



\begin{table}[t] \scriptsize
\vspace{-0.85cm}
\centering
\caption{Final set of selected patterns and their concise behavioral definitions.}
\label{table:patterns}
\begin{tabular}{p{0.25\textwidth} p{0.68\textwidth}}
\hline
\textbf{Pattern Name} & \textbf{ Definition} \\
\hline
Sequence & Enforces a strict execution order across a series of events, requiring the entire sequence to complete once initiated. \\
Deferred Choice & Provides mutually exclusive options where the execution of one alternative permanently disables all others. \\
Bounded Exists & Constrains the occurrence of an event to a parameterized threshold, enforcing a minimum ($\geq n$), maximum ($\leq n$), or exact ($= n$) execution count. \\
Explicit Initialization & Specifies a mandatory starting event that must be executed before any other activity in the process can occur. \\
Explicit Termination & Defines a specific event that immediately finishes the entire process upon its execution, preventing further activities. \\
Exists & Mandates that a designated event must occur at least once before the process can successfully finish. \\
isAbsent & Strictly prohibits the execution of a specified event throughout the entire process instance. \\
Precedes & A set-based constraint where all events in a precursor set must execute before any event in a target set is permitted. \\
LeadsTo & A set-based rule dictating that the execution of any event in a trigger set necessitates the eventual execution of all events in a target set. \\
\hline
\end{tabular}
\end{table}

\subsection{Step 2: Pattern Formalization}
\label{sec:usecase-step2}

\noindent\textit{Behavioral Specification (Step 2.1).}
For each selected pattern, we characterized the intended behavior through representative valid and invalid traces. Since the original pattern definitions are specified in natural language, this characterization required resolving ambiguities when mapping their intended semantics to DCR. We therefore selected the interpretation that best preserved the pattern's behavioral intent within DCR's semantics. For instance, since every DCR event is inherently repeatable, the chosen event in \textit{Deferred Choice} may be executed multiple times, and \textit{Sequence} must be completed once initiated, with ordering preserved across repetitions. For \textit{Explicit Initialization} and \textit{Explicit Termination}, we opted for process-level scope, meaning that the initial event precedes \emph{all} events and the terminating event finishes the \emph{entire} process. For \textit{Precedes} and \textit{LeadsTo}, we extended the original atomic definitions~\cite{elgammal2016formalizing} to sets of events. 

As an example to illustrate our behavioral specification approach, for a \textit{Sequence} over $A$, $B$, $C$, valid traces include $A \rightarrow B \rightarrow C$ and $A \rightarrow B \rightarrow C \rightarrow A \rightarrow B \rightarrow C$. Invalid traces include $A$, $A \rightarrow B$, $B \rightarrow C \rightarrow A$, $A \rightarrow C$, $A \rightarrow B \rightarrow B \rightarrow C$, and $A \rightarrow B \rightarrow C \rightarrow A$. Similar specifications were established for all nine patterns.
\vspace{0.1cm}

\begin{figure}[t]
    \centering
    \includegraphics[scale=0.094]{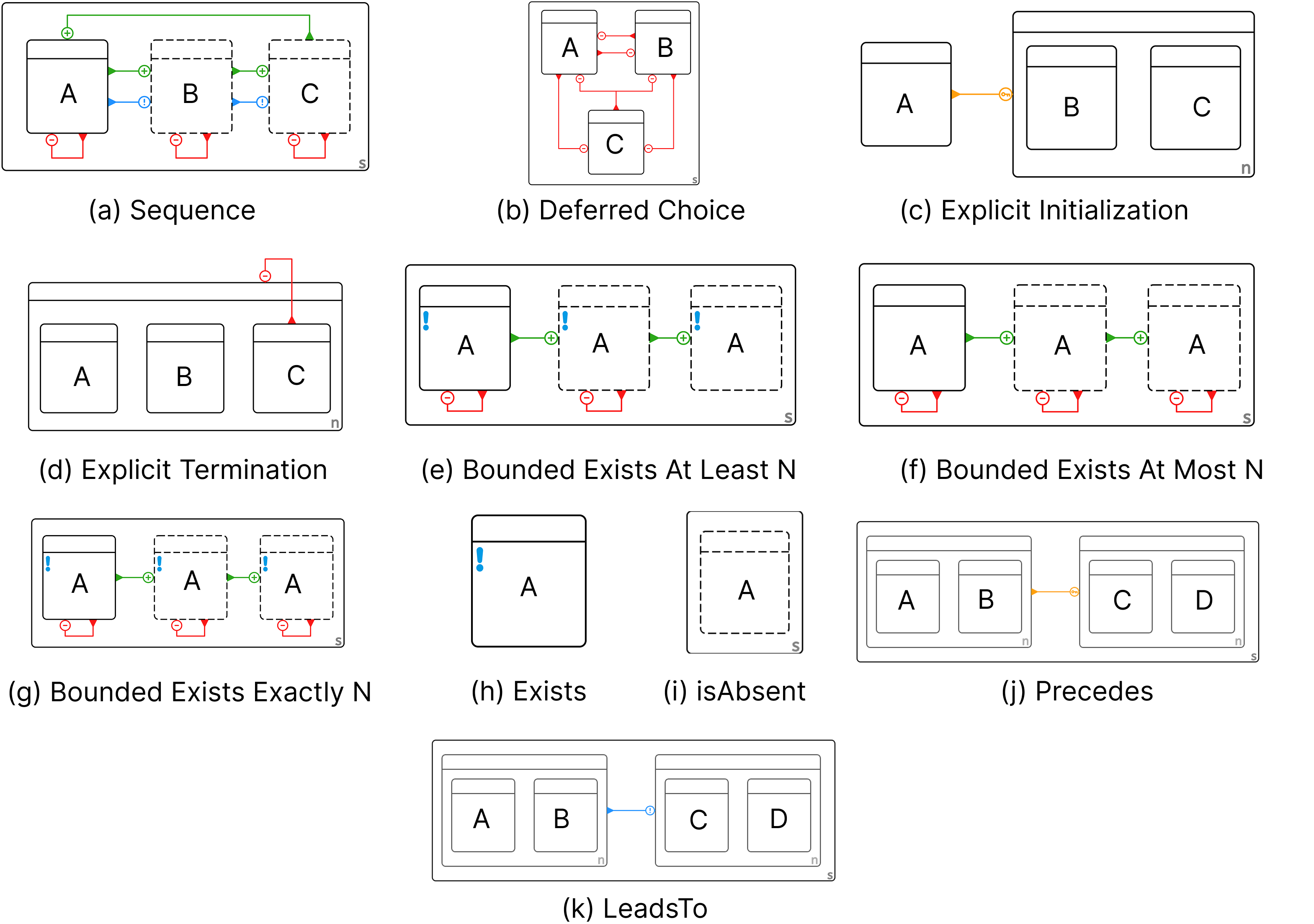} 
    \caption{DCR model fragments of the identified patterns. For the sake of illustration, we set N to 3.
    \label{fig:pattern-dcr-formalizations}}
    \vspace{-0.5cm}
\end{figure}

\noindent\textit{Behavioral Encodings (Step 2.2).}
Each pattern was then translated into a DCR graph. The encapsulation principle was a fundamental design decision for the encodings, i.e., \emph{no DCR relation may cross a pattern's boundary}. 
This prevents external interference with internal pattern behavior and ensures that each pattern preserves its intended semantics when composed.
The resulting DCR graphs for all nine patterns are shown in Fig.~\ref{fig:pattern-dcr-formalizations} (labeled from (a) to (k)). {We refer the reader to our repository for a detailed explanation of the DCR construct configuration underlying each pattern \cite{appendix}.} 

\vspace{0.1cm}

\noindent\textit{Parameterization (Step 2.3).}
\textit{Bounded Exists} is parameterized by the threshold $N \in \mathbb{N}$, which determines the number of identically labeled events, while the constraint type, i.e., at least, at most, or exactly, determines the pending markings and exclusion relations. \textit{Precedes} and \textit{LeadsTo} are parameterized by the cardinalities of their event sets $X$ and $Y$. The other patterns are similarly parameterized by their number of constituent events (cf. Fig.~\ref{fig:pattern-dcr-formalizations}).

\subsection{Step 3: Visual Notation Design}
\label{sec:usecase-step3}


\noindent\textit{Principle Selection (Step 3.1).}
Following our method's guidance, we adopted two complementary frameworks. From \textit{PoN}~\cite{moody2009physics}, we prioritized \emph{semantic transparency} and \emph{dual coding} to address interpretability. From the \textit{Semiology of Graphics}~\cite{bertin1983semiology}, we drew on its visual variables (shape, color, size, orientation, brightness, texture) to ensure distinguishability across pattern symbols.

\vspace{0.1cm}

\noindent\textit{Symbol and Label Design (Step 3.2).}
We generated five candidate \emph{symbols} per pattern, drawn from public signage, traffic signs, and BPMN notation to maximize semantic transparency and visual distinctiveness. We also generated nine candidate \emph{labels} per pattern, using synonyms and abbreviations to complement the symbols through dual coding. {The symbols and labels pairs are available in our repository as part of our Delphi survey~\cite{appendix}.}
\vspace{0.1cm}

\begin{figure}[t]
    \centering
    \vspace{-0.5cm}
    \includegraphics[scale=0.55]{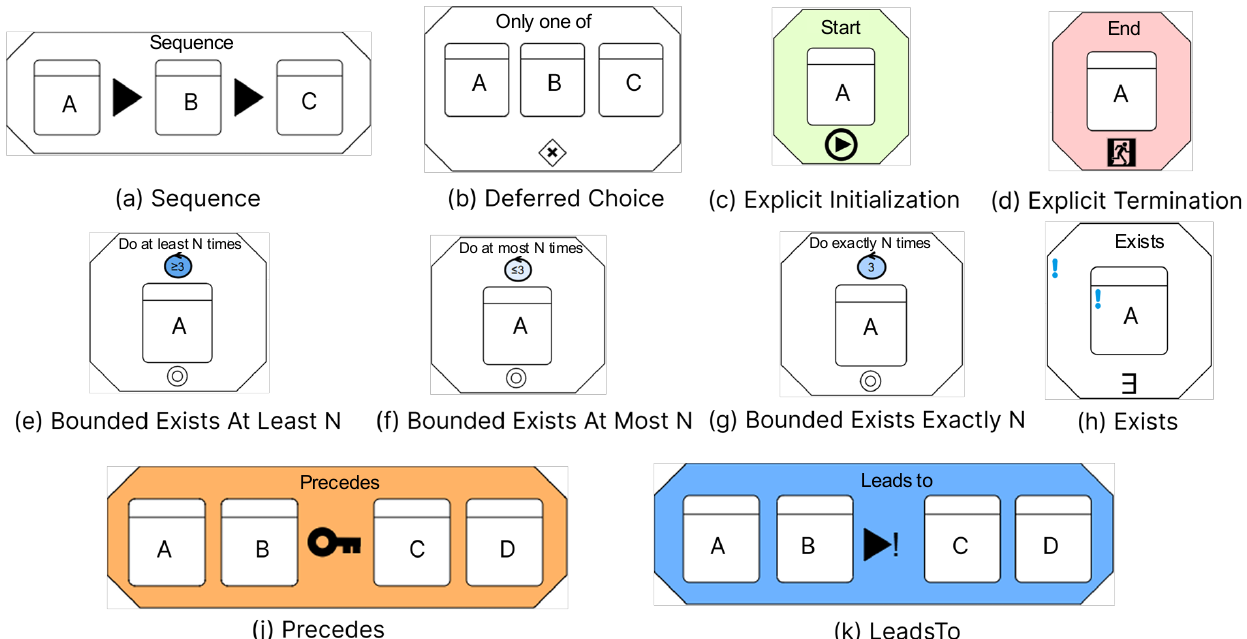} 
    \caption{DeCleaR Patterns. Note: The Bounded Exists patterns (e,f,g) include a dynamic circular bar at the bottom denoting a progress bar that advances at run time to show how many times the pattern has been executed. }
    \label{fig:declear-pattern-views}
    \vspace{-0.5cm}
\end{figure}

\noindent\textit{Panel Validation of Designs (Step 3.3).} 
To converge on final designs, we conducted a two-round Delphi survey~\cite{okoli2004delphi}, involving an associate professor (expert in DCR and conceptual modeling), 2 senior researchers (both experts in conceptual modeling; one expert in DCR), and three graduate students (with one year of training in DCR and other modeling languages). Anonymous voting mitigated authority bias. Participants ranked symbol, label, and visual-variable candidates based on interpretability and distinguishability. Aggregated responses were shared in the second round, and convergence required majority consensus on each symbol-label pair. {Our Delphi survey is available in our online repository~\cite{appendix}.} The results confirmed color and shape as dominant variables for distinguishability and converged toward the DeCleaR representations shown in Fig.~\ref{fig:declear-pattern-views}.

\subsection{Step 4: Empirical Validation}
\label{sec:usecase-step4}


\noindent\textit{Hypotheses and Comprehension Operationalization (Step~4.1).}
We formulated \textit{four null hypotheses} targeting distinct comprehension dimensions. \textbf{H\textsubscript{1}} tests {answer correctness}: \textit{DeCleaR and DCR exhibit the same error rate during decoding tasks}. {H\textsubscript{2}} and {H\textsubscript{3}} test {perceived quality} via the SEQUAL framework~\cite{krogstie2012model}: \textbf{H\textsubscript{2}} posits \textit{no significant difference in empirical quality}, while \textbf{H\textsubscript{3}} posits \textit{no significant difference in pragmatic quality.} \textbf{H\textsubscript{4}} tests {preference}: \textit{participants favor DeCleaR and DCR equally}. 
\vspace{0.1cm}

\noindent\textit{Experimental Design (Step~4.2).}
We adopted a counterbalanced within-subject design with fourteen participants familiar with DCR, recruited from university courses, BPM research groups, and industry partners\footnote{Anonymized for the reviewing process.}
. To demonstrate the abstract notation engineering method while limiting fatigue, four DeCleaR patterns were selected for a first experiment: \textit{Bounded Exists at Least N times}, \textit{Sequence}, \textit{Explicit Initialization}, and \textit{Explicit Termination}. The selection was made through anonymous co-author voting. Each pattern was embedded in an identical base model, ensuring that the notation variant was the only manipulated variable.

Our data collection combined four instruments. Decoding tasks (H\textsubscript{1}) required participants to answer with valid execution traces~\cite{abbad2023complexity}. A Likert questionnaire (H\textsubscript{2}, H\textsubscript{3}) measured empirical and pragmatic quality using items adapted from validated instruments~\cite{DEMOunderstandability}. User preference comparisons (H\textsubscript{4}) presented semantically identical DeCleaR and DCR representations side by side. Finally, semi-structured interviews explored participants' reasoning about symbols, labels, colors, shapes, pattern difficulty, and perceived complexity. {Our experiment material is available in our online repository \cite{appendix}.}
\vspace{0.1cm}

\noindent\textit{Analysis and Cross-Method Synthesis (Step~4.3).}
With fourteen participants, four patterns, and one task per pattern, we obtained 56 data points for DCR and 56 for DeCleaR. Given non-normal distributions, we tested the hypotheses using Wilcoxon signed-rank paired tests. Interview transcripts were analyzed through grounded-theory-inspired coding (involving initial coding, focused coding, and axial coding)~\cite{CodingManualQualitative}, with substantial inter-rater reliability (Cohen's~$\kappa=0.82$). {Our coding is documented in our online repository \cite{appendix}.}

H\textsubscript{1} could not be rejected ($p=0.23$), indicating no significant difference in answer correctness. H\textsubscript{2} ($p=0.013$, median~$\Delta$=+0.667), H\textsubscript{3} ($p=0.012$, median~$\Delta$=+1.5), and H\textsubscript{4} ($p<0.001$, median~$\Delta$=+1.0) were rejected in favor of DeCleaR, showing significantly improved perceived empirical quality, pragmatic quality, and preference, respectively.
The integration of qualitative and quantitative insights revealed a central \emph{perception-performance gap}. Indeed, subjective dimensions improved, while answer correctness remained unchanged. Our qualitative coding could explain this divergence. Participants found the individual DeCleaR patterns intuitive and clearly distinguishable, symbols expressive, labels useful, and their combination supportive for interpretability. However, they struggled to trace how external DCR relations interact with the patterns' behaviors within the process models, resulting in comparable error rates across notations. Our qualitative coding also yielded 42 exploratory usability insights. {For brevity, these insights are reported in our repository \cite{appendix}.}

\vspace{-0.2cm}
\section{Discussion}
\label{sec:discussion}


This paper's primary contribution is a systematic, language-agnostic method for designing visual notation abstractions. Unlike prior work proposing notation-specific visual improvements~\cite{trinh2023semantic,3DDCR}, our method provides a general framework grounded in pattern-oriented conceptual modeling~\cite{russell2006workflow,elgammal2016formalizing}, trace-based behavioral specification~\cite{Zugal2013b}, visual notation design principles~\cite{moody2009physics,bertin1983semiology}, and empirical validation guidelines~\cite{wohlin2003empirical}. The DCR use case serves as a proof of concept demonstrating the method's applicability.

In this use case, applying our method to DCR resulted in DeCleaR, which participants rated higher on empirical and pragmatic quality, and preferred it over standard DCR representations. Answer correctness, however, showed no significant difference between conditions. This suggests that DeCleaR improves user experience without degrading accuracy, while the complexity of surrounding DCR relations may still dominate task difficulty. These insights are consistent with prior work on declarative model comprehension~\cite{reijers2013declarative,abbad2020exploring}. Our results can guide the next iteration of notation engineering. For DeCleaR, this means preserving the pattern-level abstractions while improving how their interaction with surrounding DCR relations is represented and supported.
\vspace{0.1cm}

\noindent\textit{Implications.} For \emph{notation designers}, our method provides a structured, theory-grounded direction to extend existing languages with pattern-based visual abstractions. For \emph{tool builders}, the formalization step produces specifications that can be implemented as transformation rules or visual overlays. For \emph{practitioners and educators}, pattern-based abstractions bridge domain vocabulary and formal notation, enabling faster onboarding and reducing the expertise barrier that may limit declarative modeling adoption~\cite{abbad2020exploring}. More broadly, the method opens a design space for \emph{domain-specific modeling profiles}, where communities can systematically develop and standardize patterns relevant to their practice.
\vspace{0.1cm}

\noindent\textit{Limitations.} 
The method was instantiated on one language, although it was designed to be language-agnostic. To avoid overfitting, we grounded each step in language-independent foundations that have been applied to several conceptual modeling languages. Still, this creates a potential threat to external validity, and future work should apply the method to other languages. A second limitation concerns the scope of the DCR instantiation, which involved a limited participant group and covered few patterns. These decisions align with the paper's objective, namely demonstrating that visual notation abstractions \emph{can} be systematically designed and \emph{do} yield measurable benefits. Exhaustive pattern coverage and large-scale validation remain necessary future steps.








\section{Conclusion and Future Work}
\label{sec:conclusion}



In this work, we presented a method for designing visual notation abstractions that replace complex configurations of low-level constructs. Grounded in established theoretical foundations, the method comprises four replicable steps. Its instantiation on DCR graphs produced DeCleaR and showed positive effects on empirical quality, pragmatic quality, and user preference.

Future work should apply the method to other languages to further validate its generalizability. For DeCleaR, broader evaluations should include more patterns and larger participant groups. Eye-tracking studies could further examine how abstractions affect practitioners’ cognitive processes, while tool integration would enable evaluation at scale, extending to modeling and maintenance tasks.

\paragraph{Acknowledgments} This work has been supported by the grant ``Center for Digital CompliancE (DICE)'' (VIL57420) from VILLUM FONDEN, and by the Innovation Foundation project ``Explainable Hybrid-AI for Computational Law and Accurate Legal Chatbots'' 4355-00018B XHAILe, and Predictive and Prescriptive Process Analytics for Industry 4.0 (P3AI4, grant 4105-00045B )

 \bibliographystyle{splncs04}
 
 \bibliography{literature}
\end{document}